# Revival of Strain Susceptibilities: Magnetostrictive Coefficient and Thermal-Expansion Coefficient

Yisheng Chai[1*]

[1]*Low Temperature Physics Laboratory, Center of Quantum Materials and Devices and College of Physics, Chongqing University, Chongqing 401331, China*
*Corresponding author (emails: yschai@cqu.edu.cn);

In thermodynamics, volume is an essential extensive variable. Strain—line, area, or volume change—therefore offers a direct window into correlated quantum matter: tiny length changes $\Delta L$ track how the lattice responds when state variables such as magnetic field $H$ and/or temperature $T$ are varied, revealing phases, transitions, and dynamics. Direct, high-precision strain measurements are already difficult; their susceptibilities are harder still. Very recently, several direct techniques have made vital progress on two key quantities: the magnetostrictive coefficient $d\lambda/dH$ (often denoted $q_{ijk}$ or $d_{ij}$ in the magnetostriction literatures) [1-5], and the linear thermal-expansion coefficient $\alpha = d\lambda/dT$ [6,7]. Considering these two strain susceptibilities together—they are fundamental and complementary—clarifies why these thermodynamic properties merit renewed attention. It also highlights a real asymmetry: our composite magnetoelectric (ME) technique (Fig. 1a) has turned the ac magnetostrictive coefficient $(d\lambda/dH)_{ac}$ into a fast, high-contrast observable under extreme conditions (e.g., down to 1.7 K in temperature and up to 33 T in magnetic field at the High Magnetic Field Laboratory of the Chinese Academy of Sciences, CHMFL), whereas truly direct, dynamic $\alpha$ measurements remain comparatively sparse and rely on more complex piezoelectric or optical approaches; many studies still obtain $\alpha$ by differentiating precise $\lambda(T)$ from dc dilatometry [8]. The calculated $d\lambda/dH$ and $\alpha$ in Ref. 8 has the order of magnitude of several $10^{-5}$/T and $10^{-5}$/K, respectively, with a resolution of about $10^{-6}$/T or $10^{-6}$/K.

The theoretical basis is thermodynamic. For small strain $\varepsilon_{ij}$ (a second-order tensor) and away from structural instabilities, the incremental lattice response to stress $\sigma$, field $H$, and temperature $T$ is:

$$d\varepsilon_{ij} = s_{ijkl}d\sigma_{kl} + q_{ijk}dH_k + \alpha_{ij}dT,$$

where $s_{ijkl}$ is elastic compliance tensor, $q_{ijk}\equiv\partial\varepsilon_{ij}/\partial H_k$ the magnetostrictive susceptibility tensor, and $\alpha_{ij}\equiv\partial\varepsilon_{ij}/\partial T$ the thermal-expansion tensor (all at fixed $T$, $H$, $\sigma$). Projecting onto a measurement axis $n$ gives $d\lambda/dH_k = n_i q_{ijk} n_j$ and $\alpha_n = n_i \alpha_{ij} n_j$. Under ac drive, both become complex response functions: $(d\lambda/dH)_{ac}=d\lambda'/dH+id\lambda''/dH$ and $\alpha_{ac}=\alpha'+i\alpha''$; with the in-phase channel reflecting reversible elastic response and the out-of-phase channel quantifying dissipation (domain-wall friction, thermoelastic damping, superconducting vortex motion, etc.). Maxwell identities derived from the Gibbs free-energy density $g(\sigma, H, T)$ tie these strain susceptibilities to their conjugate variables,

$$\partial\varepsilon_{ij}/\partial H_k|_{\sigma,T} = -\partial M_k/\partial\sigma_{ij}\big|_{H,T}, \quad \partial\varepsilon_{ij}/\partial T|_{\sigma,H} = -\partial S/\partial\sigma_{ij}\big|_{H,T},$$

where $M$ and $S$ are magnetization and entropy, respectively, explaining the sensitivity of $d\lambda/dH$ to spin correlations (via stress derivatives of magnetization) and of $\alpha$ to entropy changes near phase boundaries.

To measure $(d\lambda/dH)_{ac}$ directly, we use a composite-ME configuration introduced in multiferroics around 2000 [9]. A magnetic specimen is bonded to a piezoelectric single crystal, forming a laminate that converts tiny ac magnetostrictive strain into an electrical signal proportional to $(d\lambda/dH)_{ac}$. Lock-in detection reads the regulated real and imaginary parts $d\lambda'/dH$ and $d\lambda''/dH$ directly (Fig. 1a). Three practical features make this method attractive: (i) easy, quick mounting, vibration-tolerant measurement that works in essentially any cryostat and high-field environment; (ii) narrowband, lock-in readout that reaches sub-nanostrain-equivalent sensitivity and supports fast field/temperature sweeps (on the order of a second per point) for dense $H$–$T$ phase diagram mapping; and (iii) a signal proportional to the field derivative of strain itself, so reversible in-phase changes track elastic evolution, while loss peaks and hysteresis mark kinetic or first-order processes. A decisive detail is the piezo layer: using $0.7Pb(Mg_{1/3}Nb_{2/3})O_3$-$0.3PbTiO_3$ (PMN-PT) near the morphotropic phase boundary (~30% PT) places the crystal in the rhombohedral–tetragonal region where phase coexistence and polarization rotation yield exceptionally large piezoelectric coefficients[10]; this enables wide $T$–$H$ operation, and the boundary persists essentially down to 0 K with negligible field dependence. Absolute calibration of $d\lambda/dH$ from piezo-based signals is intrinsically difficult because the electromechanical gain depends on bonding mechanics and on a temperature-dependent piezoelectric coefficient. We therefore present normalized $(d\lambda/dH)_{ac}$ and reserve absolute $(d\lambda/dH)_{dc}$ for dc capacitance dilatometry [11].

In ferro- and antiferromagnetics, $d\lambda'/dH(H,T)$ traces magnetic ordering, spin-reorientation and saturation lines and canted phases (Fig. 1b), while $d\lambda''/dH$ separates first-order boundaries—hysteretic, with pronounced negative sign (Fig. 1c)—from second-order ones that evolve reversibly without loss[1]. Moreover, $(d\lambda/dH)_{ac}$ is a targeted probe of criticality. At a tricritical point—where a first-order line meets a second-order line—$d\lambda'/dH$ often shows a local maximum (Fig. 1c) [4]. Similarly, $d\lambda'/dH$ is enhanced at triple points in layered van der Waals ferromagnets with strong spin–spin correlations [1]. In chiral magnets, the method reveals skyrmion phases with exceptional clarity: steps or kinks in the real channel mark entry/exit, while dissipation peaks at the boundaries signal a liquid-like skyrmion regime or phase coexistence; single crystals typically present a compact skyrmion-lattice pocket [2].

In clean metals, the same strain susceptibility becomes a powerful quantum-oscillation probe. Because the measured quantity is the field derivative of $\lambda(H)$, oscillatory components are emphasized and onset fields are often lower than in magnetization, resistivity, thermopower, or DC dilatometry (Fig. 1e). Under moderate fields one can resolve six fundamental orbits in ZrSiS (Fig. 1f); temperature-dependent amplitudes follow the Lifshitz–Kosevich form, enabling effective-mass extraction. High-frequency components in the several-kT range can appear at just 13 T and 2 K (Fig. 1g), rarely seen by other probes on the same sample [3].

In type-II superconductors, the composite-ME technique reveals a significant magnetostrictive response. Using Nb as a case study, we demonstrate that the AC response $(d\lambda/dH)_{ac}$ is distinct from the DC value $(d\lambda/dH)_{dc}$ obtained by differentiating $\lambda(H)$ from DC dilatometry (Figs. 1h and 1i); this establishes the dynamic magnetostrictive effect as a unique phenomenon in type-II superconductors. Across several representative materials, the real component d$\lambda'$/d$H$ scales linearly with vortex density within the vortex lattice phase (Figs. 1j and 1k), while the imaginary component d$\lambda''$/d$H$ remains negligible. This implies that the AC slope reflects a collective elastic mode of the pinned vortex array in the small-amplitude limit, whereas (d$\lambda$/d$H$)$_{dc}$ captures the quasi-static magnetoelastic background alongside irreversible vortex entry/exit. Upon entering the vortex liquid phase at Hirr, dissipative motion causes d$\lambda''$/d$H$ to emerge and peak. At higher fields, where both d$\lambda'$/d$H$ and d$\lambda''$/d$H$ vanish, the vortex density reaches zero, defining the upper critical field $H_{c2}$ (Fig. 1i) [5].

Thermal expansion completes the picture. Direct, dynamic α—apply a small temperature modulation Δ$T$ and lock-in detect d$\lambda$/d$T$—avoids numerical differentiation of $\lambda(T)$ and offers the same phase sensitivity that makes (d$\lambda$/d$H$)$_{ac}$ so diagnostic. A temperature-modulated (piezobender-based) dilatometer operates at cryogenic temperatures with high resolution on correlated metals and superconductors [6], while a high-resolution all-fiber Michelson interferometer measures α without piezo gain and has resolved antiferromagnetic transitions in iron pnictides [7]. Absolute α from piezo pickups remains very difficult for the same transduction-gain reasons, so dynamic α reports are fewer and often system-specific.

Looking forward, the quickest way to symmetry is practical: make α as convenient as composite-ME measured d$\lambda$/d$H$ by integrating micro-heaters and thermometry into simple holders (with minimal, simple calibration) while converting $\lambda$ into electrical or optical signals. With those steps, the zoo of strain susceptibilities can become an everyday toolkit for mapping $H$–$T$ phase diagrams and dynamics across correlated, magnetic, and superconducting materials.

**Conflict of interest**
The authors declared no conflict of interest.

**Figure 1** | (a) Composite-ME laminate with ac magnetic drive and lock-in detection; the measured voltage is proportional to d$\lambda'$/dH+id$\lambda''$/dH. (b) The magnetostriction coefficient of Kitaev spin liquid candidate, Na$_3$Co$_2$SbO$_6$ measured as a function of magnetic field captured at selected temperatures (H//a). (c) and(d) Comparison of the real and imaginary parts of the ac magnetostriction coefficient and ac magnetic susceptibility recorded at 2 K [4]. (e) Background-subtracted field-dependent oscillations of $V_x$ at selected temperatures with B//c for Dirac nodal-line semimetal ZrSiS. (f) FFT spectrum (5–1000 T) over a magnetic field range of 1–13 T. (g) FFT spectrum (7.5–11.5 kT) across a magnetic field range of 1–13 T [3]. (c) Magnetic field dependence of (h), magnetization M, (i), Real and imaginary components of the ac magnetostrictive coefficient of polycrystalline Nb superconductor. (j) Resistance and (k) d$\lambda'$/dH under various magnetic fields [5].

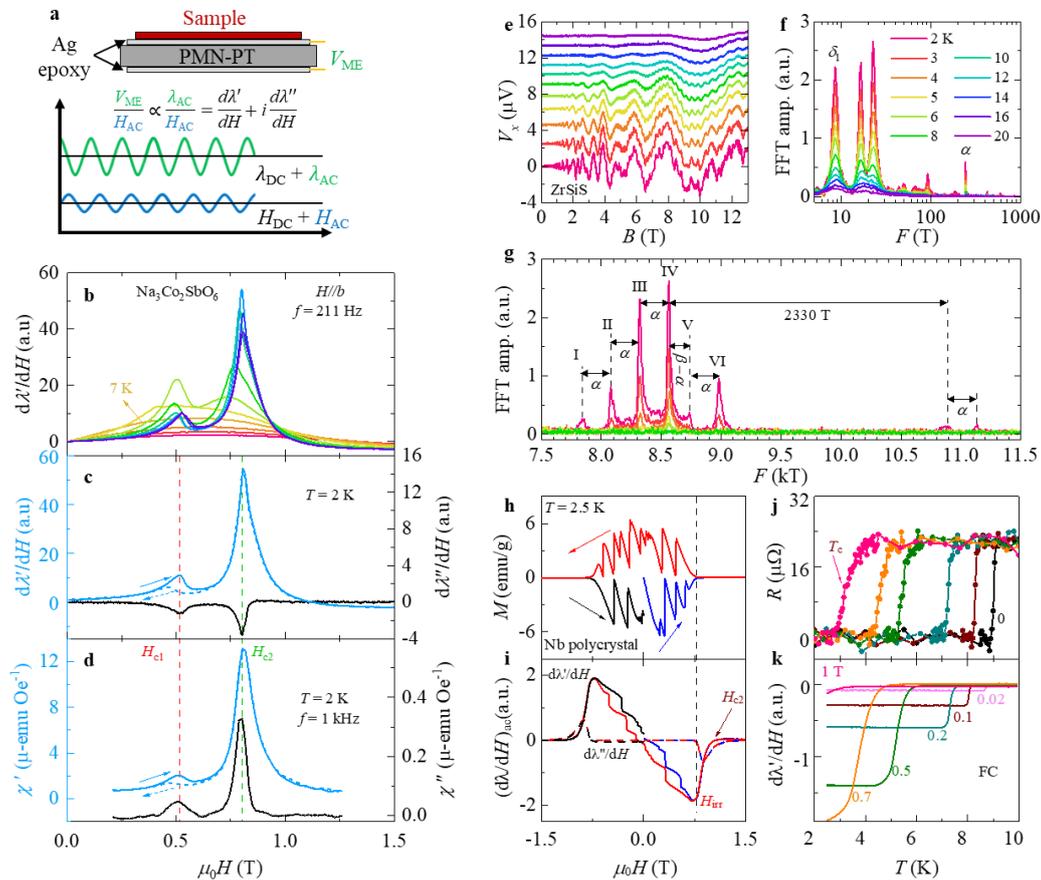